\documentclass{article}

\usepackage{arxiv}

\usepackage[utf8]{inputenc} 
\usepackage[T1]{fontenc}    
\usepackage{hyperref}       
\usepackage{url}            
\usepackage{booktabs}       
\usepackage{amsfonts}       
\usepackage{nicefrac}       
\usepackage{microtype}      
\usepackage{lipsum}		
\usepackage{graphicx}
\usepackage[numbers]{natbib}
\usepackage{doi}
\usepackage{comment}
\usepackage{xcolor}
\usepackage{multicol} 
\usepackage{cancel}
\usepackage{comment}

\title{Content Quality vs. Attention Allocation: An LLM-Based Case Study in a Peer-to-Peer \\ Mental Health Network}

\author{Teng Ye \\
	University of Minnesota \\
	\texttt{tengye@umn.edu} \\
	\And
        Hanson Yan\\
	MobLab \\
	\texttt{hanson.yan@moblab.com} \\
        \And 
        Xuhuan Huang \\
        MobLab \\
        \texttt{jason.huang@moblab.com} \\
        \And 
        Connor Grogan \\
        MobLab \\
        \texttt{connor.grogan@moblab.com} \\
        \And 
        Walter Yuan \\
        MobLab \\
        \texttt{walter.yuan@moblab.com} \\
        \And 
        Qiaozhu Mei \\
        University of Michigan \\
        \texttt{qmei@umich.edu} \\
        \And 
        Matthew O. Jackson \\
        Stanford University \\
        \texttt{jacksonm@stanford.edu} \\
}

\date{}

\renewcommand{\headeright}{}
\renewcommand{\shorttitle}{Content Quality vs. Attention Allocation: An LLM-Based Case Study in Peer-to-peer Mental Health Networks}

\begin{document}
\maketitle

\begin{abstract}

With the rise of social media and peer-to-peer networks, users increasingly rely on crowdsourced responses for information and assistance. However, the mechanisms used to rank and promote responses often prioritize and end up biasing in favor of timeliness over quality, which may result in suboptimal support for help-seekers. 
We analyze millions of responses to mental health-related posts, utilizing large language models (LLMs) to assess the multi-dimensional quality of content, including relevance, empathy, and cultural alignment, among other aspects. Our findings reveal a mismatch between content quality and attention allocation: earlier responses---despite being relatively lower in quality—--receive disproportionately high fractions of upvotes and visibility due to platform ranking algorithms. We demonstrate that the quality of the top-ranked responses could be improved by up to 39 percent, and even the simplest re-ranking strategy could significantly improve the quality of top responses, highlighting the need for more nuanced ranking mechanisms that prioritize both timeliness and content quality, especially emotional engagement in online mental health communities.

\end{abstract}

\keywords{Large language models \and content quality \and online help-seeking \and mental health}

\section{Introduction}

With the advent of online-forums and platforms, there has been a deluge of user-generated content.  Topics and threads often receive comments and contributions from thousands of users.  This includes help forums, reviews of products, discussion threads, user generated videos, etc.  Given the huge variation in the quality of such comments and contributions, platforms have to prioritize what they show users in order to present the most relevant content.  To do so most platforms have users themselves rate and vote on comments and content. 
These ranking systems are usually built upon techniques in the field of information retrieval, guided by the probabilistic ranking principle \cite{robertson1977probability}. Ideally, this allows users to focus on the most relevant and valuable content. Similarly, content-providers can prioritize topics and areas that have not yet received high-quality responses, where their expertise can have the greatest impact. These ranking mechanisms typically rely on a combination of factors, including platform-determined quality assessments, crowdsourced feedback such as `likes,' `thumbs-up,' and `upvotes,' as well as factors like timeliness or the credentials of the responders.

Despite the wide use of these ranking mechanisms, their ability to correctly rank the true quality and usefulness of content has rarely been comprehensively analyzed. This gap exists largely due to the lack of scalable methods for assessing the quality and utility of content. The challenge is particularly pronounced in contexts such as seeking health-related advices, where the quality of a response is multi-dimensional. Factors range from the relevance of the response to the original question and the accuracy of the answer, to more personal and emotional aspects such as empathy and whether the response aligns with the cultural context of the help-seeker. These dimensions are reported to be critical for the success of online help-seeking communities (e.g., \cite{wang2012stay}).
While information retrieval (IR) methods excel at detecting the topical relevance between a document and a query, it is far more challenging to identify the nuanced emotional aspects of a post using traditional IR techniques or lexical-based approaches. In the existing literature, these emotional and personal dimensions often require human annotations (e.g., \cite{ouyang2015modeling,wang2015eliciting}), which do not scale effectively. Alternatively, machine learning classifiers trained on human assessments are used, but they face difficulties in generalizing across different aspects, communities, and platforms.

Large language models (LLMs) provide an effective and scalable solution for measuring these nuanced aspects and assessing content quality that is highly aligned with humans \cite{le2023uncovering, lin2023llm, wang2023automated}. This development presents unprecedented opportunities to analyze the multi-dimensional quality of posts at scale and to explore the interplay between content quality and attention allocation mechanisms in online help-seeking communities. In this study, we take the initiative to evaluate the quality of millions of responses to nearly one hundred thousand mental health-related posts, uncovering both the strengths and limitations of the ranking mechanisms employed by the social media platform under investigation.

We select mental health support networks for this case study for multiple reasons. Mental health is a global issue, with an urgent need for effective treatments and support strategies. In the United States alone, over 60 million adults experience mental illness each year, and one-third of young adults received treatments for mental health conditions
\cite{reinert2024state,hhs2023nsduh}. Globally, it is estimated that one in seven adolescents suffers from mental health disorders \cite{who2024adolescent}. The youth situation has worsened since the COVID-19 pandemic. A recent report shows that in 2023, 20.17\% of U.S. teens aged 12-17 experienced a major depressive episode, with 15\% of them experiencing severe impairment \cite{reinert2024state}.

Despite its prevalence, the diagnosis and treatment of mental health issues are often constrained by cultural and personal nuances. Unlike physical health, mental health is heavily stigmatized in many societies, resulting in delays in seeking treatment and limited support systems \cite{jett2019mental}. Stigma, cultural barriers, and lack of awareness prevent individuals from accessing professional help early on, exacerbating the severity of mental health problems \cite{henderson2013mental}. Access to mental health care is particularly limited in low-resource settings, with the WHO estimating more than 75\% of people with severe mental disorders in low- and middle-income countries receive no treatment \cite{world2021comprehensive}. Even in high-income countries, disparities persist in accessing mental health services, particularly in underserved areas.

As an alternative to professional counselors, many individuals turn to social networks, posting help-seeking content on platforms like Twitter, Quora, and Reddit. In response, these platforms are creating dedicated communities and features to facilitate help-seeking for mental health concerns. Communities such as subreddits or Facebook groups bring together a diverse range of help providers, including healthcare professionals, individuals who have experienced similar challenges, and others with varying levels of knowledge about mental health. The responses not only provide direct support to the help-seeker but also benefit a broader audience who face similar issues.  Leveraging large language models, our work provides an initial assessment of the multi-dimensional quality of the peer responses in these mental health networks and reveals how effectively the platform routes the users' attention to high quality responses.

\section{Method}

We collect posts from Reddit that are published in its sub-communities that facilitate advice or help seeking related to common mental health disorders. For 10 common mental health disorders, we select one relevant subreddit with the largest number of members as of December 2023, and our final dataset consists of 11 subreddits. 
Note that all selected subreddits have deployed \textit{flairs}, a tagging system organized by subreddit moderators. 
\footnote{Approximately, more than half of posts in each of our selected subreddits are tagged with at least one flair.}. We restrict our analysis to posts that are flaired, as tagging a post may have a confounding effect on the attention of audience (e.g., \cite{nam2014informational}). For each post, we collect all comments that directly replied to the main post, excluding posts with fewer than 10 direct comments. In total, our dataset contains 96,123 Reddit posts and more than 3,156,000 direct comments dated between June 2005 and December 2023. For each post and comment, we record the timestamp of when it was posted on Reddit, as well as the number of upvotes the comment had received by December 2023. 

For each of the 3,156,070 comments, we assess its quality across multiple dimensions using OpenAI’s ChatGPT-4, a widely used commercial large language model. Specifically, we instruct the model to evaluate each comment according to seven criteria, each rated on a scale of 1 to 10: relevance to the original post (\textit{topical alignment}), clarity of language (\textit{lexical precision}), use of \textit{empathy expression}(s), tone of encouragement (\textit{encouragement level}), provision of \textit{actionable suggestion}(s), presence of \textit{personal resonance}, and adherence to the cultural norms of the post’s author (\textit{cultural alignment}). Finally, the seven ratings are averaged to generate a composite quality score for each comment.

\begin{figure*}[t]
  \centering
  \vskip -10pt
  \includegraphics[width=\textwidth]{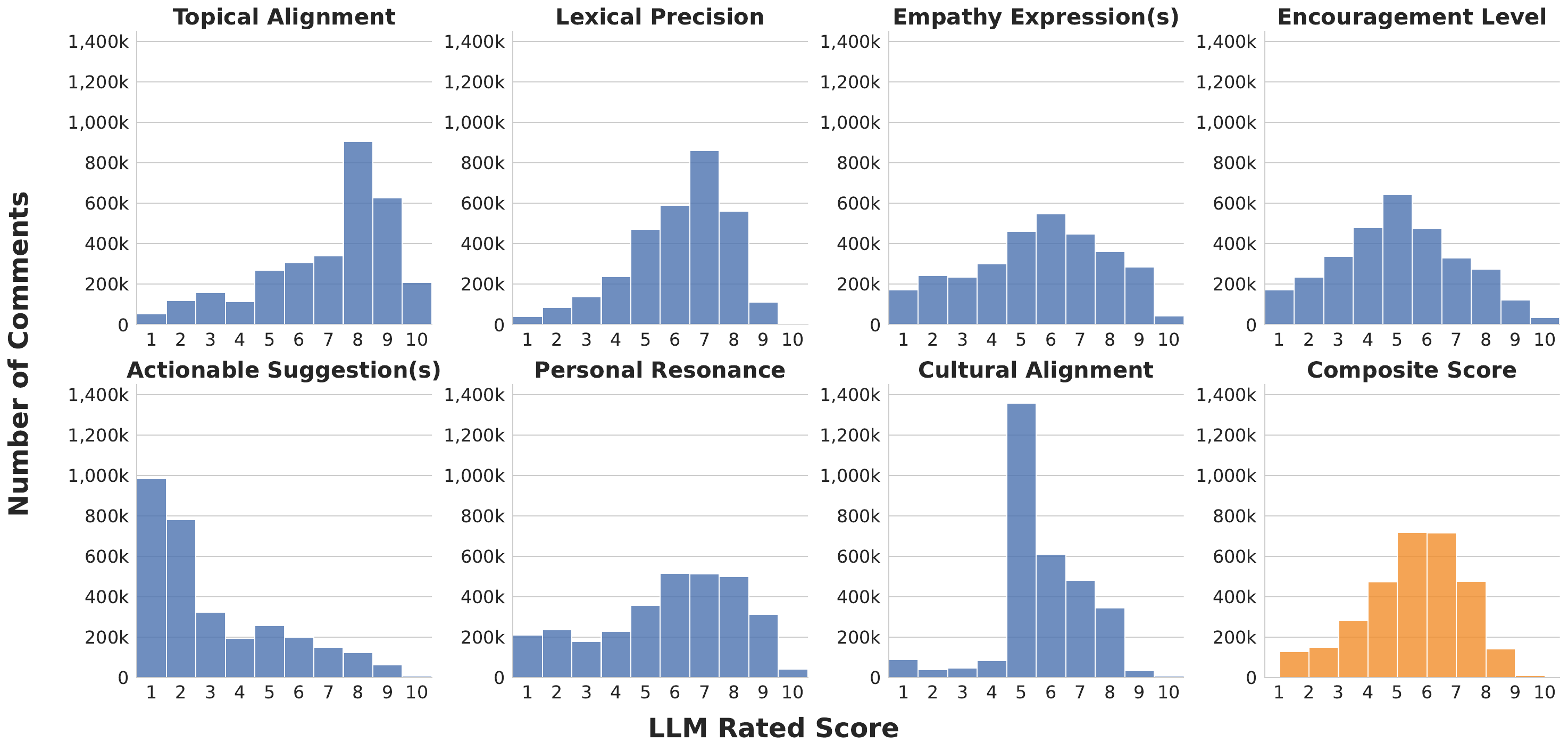}
  \caption{Distributions of the composite quality score and per-aspect scores show distinct patterns. The composite score follows a bell-shaped distribution, peaking between 5 and 6. Most per-aspect distributions are more skewed. There is a noticeable proportion of comments with low \textit{personal resonance} or low \textit{empathy expression}. Overall, the topical aspects of comment quality (\textit{topical alignment} and \textit{lexical precision}) are higher rated than emotional aspects (\textit{empathy expressions}, \textit{encouragement level}, and \textit{personal resonance}). The topical alignment score is centered around 8, while most comments lack actionable suggestions. }
  \label{fig:score-dist}
\end{figure*}

In Figure~\ref{fig:score-dist}, the composite quality score, which averages across seven aspects, exhibits a bell-shaped distribution peaking between 5 and 6, offering a clear categorization of comment quality. Most per-aspect scores are single-peaked and either left- or right-skewed. Notably, the majority of comments receive a topical alignment score of 8, indicating that these comments are generally relevant to the help-seeking inquiry.  A high average is observed in lexical precision scores. The actionable suggestion scores are skewed toward the lower end (1 and 2), indicating that most comments do not provide actionable advice. A noticeable proportion of comments are scored low in \textit{personal resonance} or \textit{empathy expressions}, suggesting that they offer minimal connection to personal experiences or empathetic expressions. 

\section{Results}

Ranking is a commonly used mechanism to direct users’ attention to a long list of items, assuming that users examine the ranked items in order and stop when they find a desirable one. This approach is guided by the Probabilistic Ranking Principle~\cite{robertson1977probability}, which suggests that items should be ranked based on the probability of their relevance to the user’s information need. On social media platforms, comments are often ranked by the number of reader votes (e.g., upvotes on Reddit \footnote{Reddit offers two upvote-based ranking mechanisms called `Top' (ranking by upvotes) and  `Best' (rank by upvote ratio), 
and the results of the two rankings are often times very similar. }), based on the assumption that each upvote serves as a crowdsourced endorsement of the comment's quality. However, this endorsement is often biased by the timing of the comment’s visibility and its position on the page.

Do top-ranked comments under a post accurately necessarily have the highest quality? We calculate the average quality of top-ranked comments by the platform and compare it with a hypothetical scenario where comments are ranked by their composite quality scores. This hypothetical ranking serves as an `upper bound' for the quality of the top-ranked comments as rated by the LLM, or how effectively user attention could be directed towards high-quality responses by the LLM. As shown in Table~\ref{table:quality_top_ranked_comments}, the quality of the highest-ranked comment could improve by up to 39\% compared to the current ranking algorithm used by the platform if the comments were ordered by their LLM-rated composite quality scores. As user attention shifts down to the top 3, 5, and 10 comments, the average quality of these comments decreases, but the hypothetical ranking could still improve over the platform’s ranking by 30\%, 25\%, and 16\%, respectively.

~\\

\begin{table}[htbp]
\centering
\vskip -40pt
\caption{Quality of top-ranked comments can be significantly improved by ranking with LLM-rated quality scores. }
\begin{tabular}{cccc}
\hline
\hline
\\[-10pt]
 & \multicolumn{3}{c}{\textbf{Average Comment Quality (Composite Score)}} \\ 
 &\multicolumn{3}{c}{\small\textbf{(Mean $\pm$ Standard Deviation)}}\\
\\[-10pt]
\cline{2-4}
\\[-10pt]
\textbf{Top-K} & \textbf{Ranked by the Platform} & \textbf{Ranked by Composite Score} & \textbf{Improvement} \\ 
\\[-10pt]\hline
\\[-10pt]
1 & $5.63 \pm 1.85$ & $7.83 \pm 0.86$ & 39.1\%\\ 
\\[-10pt]
3 & $5.74 \pm 1.22$ & $7.45 \pm 0.87$ & 30.0\%\\
\\[-10pt]
5 & $5.72 \pm 1.08$ & $7.17 \pm 0.91$ & 25.3\%\\ 
\\[-10pt]
10 & $5.62 \pm 0.94$ & $6.54 \pm 1.07$ & 16.4\%\\ 
\\[-10pt]
\hline\hline
\end{tabular}
\label{table:quality_top_ranked_comments}
\end{table}

\begin{figure*}[htbp]
  \centering
  \includegraphics[width=\textwidth]{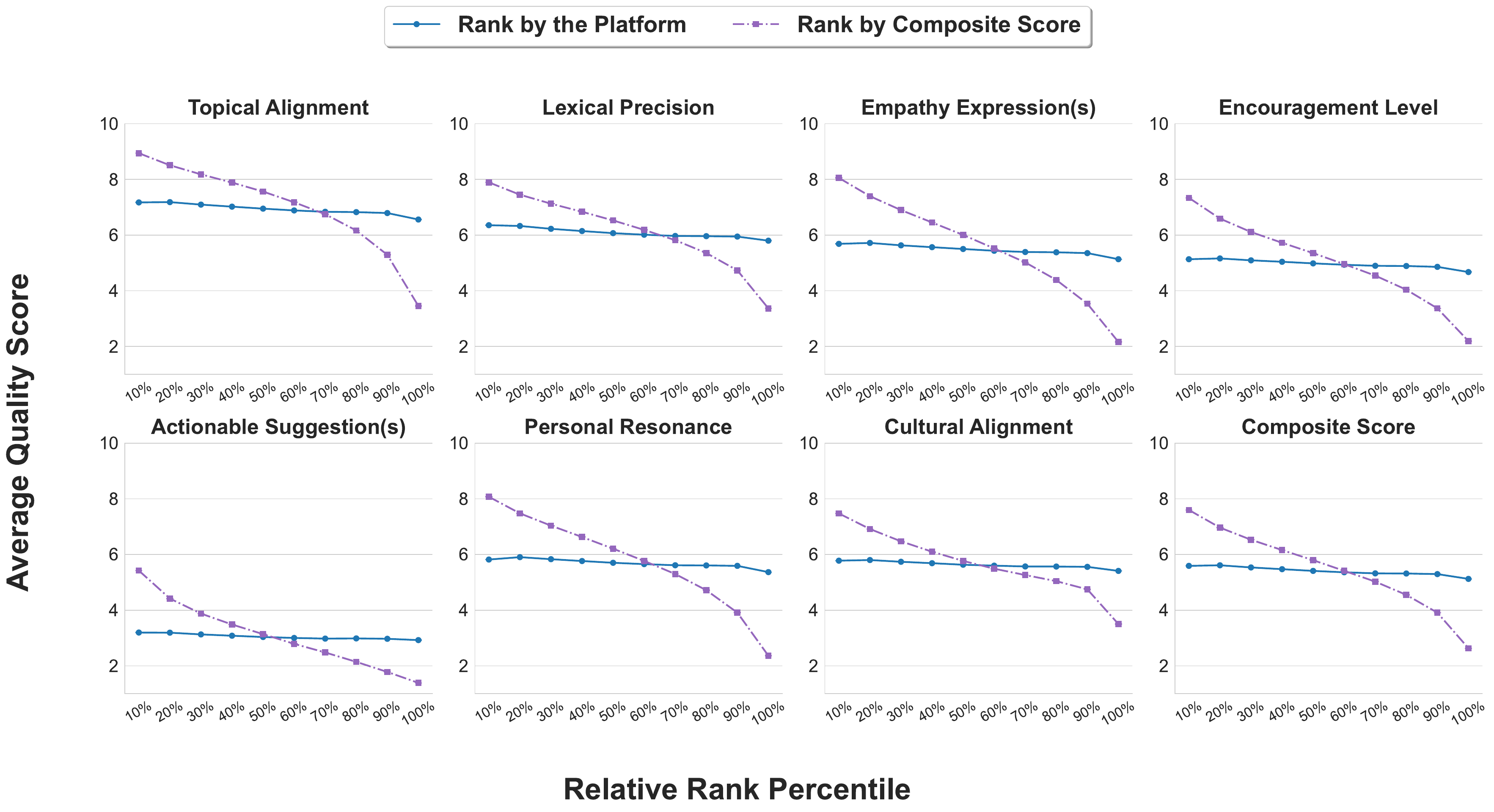}
  \caption{Average quality score at each decile of platform ranking vs. quality-based ranking. Quality of comments generally declines with the percentile of their platform ranking under the same post. However, the top 10\% of comments under each post, as ranked by the platform, have significantly lower overall quality (i.e., \textit{composite score}, $p < .001$)  and per-aspect quality for \textit{topical alignment} ($p < .10$), \textit{empathy expression} ($p < .001$), \textit{encouragement level} ($p < .001$) \textit{personal resonance} ($p < .001$), \textit{cultural alignment }($p < .001$)) than the next 10\%, particularly in non-topical and emotional aspects. Comments ranked as top 10\% have significantly higher lexical precision ($p < .001$) than those in the next decile. Actionable suggestion(s) doesn't present a significant difference between comments in the first and second decile ($p = 0.39$). In the hypothetical setting, if the comments under a post are ranked by their composite quality scores, the average quality of the top-ranked comments can be significantly boosted in all seven aspects, particularly in emotional aspects (\textit{empathy expression}, \textit{encouragement level}, and \textit{personal resonance}). }
  \label{fig:quality-over-ranking}
\end{figure*}

It is surprising to note that the average quality of the very top comment is not as high as that of top 3 (or top 5) comments, and neither is noticeably higher than the median of all comments (see Figure~\ref{fig:score-dist}). This raises the question of whether the platform ranking, or the number of upvotes, is correlated with the content quality at all. To investigate this, we bin the comments based on their percentile rank relative to all responses to the corresponding post, accounting for the varying number of comments each post receives. As shown in Figure ~\ref{fig:quality-over-ranking}, comment quality generally declines with increasing percentile rank, indicating an overall correlation between quality measures and the number of upvotes a comment receives. Pearson correlation analyses show weak negative relationships between comment upvote rank percentiles and comment quality metrics, including \textit{topical alignment} ($r=-0.08, p < .001$), \textit{lexical precision} ($r=-0.10, p < .001$), \textit{empathy expression(s)} ($r=-0.07, p < .001$), \textit{encouragement level} ($r=-0.06, p < .001$), \textit{actionable suggestions(s)} ($r=-0.04, p < .001$),
\textit{personal resonance} ($r=-0.06, p < .001$),
\textit{cultural alignment} ($r=-0.07, p < .001$), and \textit{composite score} ($r=-0.08, p < .001$). However, we again observe a surprising pattern emerging at the top of the ranking: the quality of comments in the top 10$^{th}$  percentile is lower than those in the next 10$^{th}$ percentile, contradicting the assumption of the Probabilistic Ranking Principle. The crowdsourced ratings utilized by the platform, the number of upvotes, fail to allocate reader's attention to responses with the highest quality. Hypothetically, if the comments were ranked by the LLM-rated composite quality score, the  average scores of top-ranked comments in all seven aspects are significantly higher and decrease monotonically with the rank.  This improvement is particularly salient in emotion-related aspects (\textit{empathy expression}, \textit{encouragement level}, and \textit{personal resonance}).

Why does this mismatch occur primarily among highly ranked comments? There are several possible explanations. The most likely hypothesis is that these highly upvoted comments may have been boosted by factors beyond their intrinsic quality. One such factor is timing—how quickly a comment is posted after the original post. Early comments have more visibility and are more likely to receive upvotes simply due to their position, regardless of quality.


\begin{figure}[t]
  \centering
  \vskip -10pt
  \includegraphics[width=\linewidth]{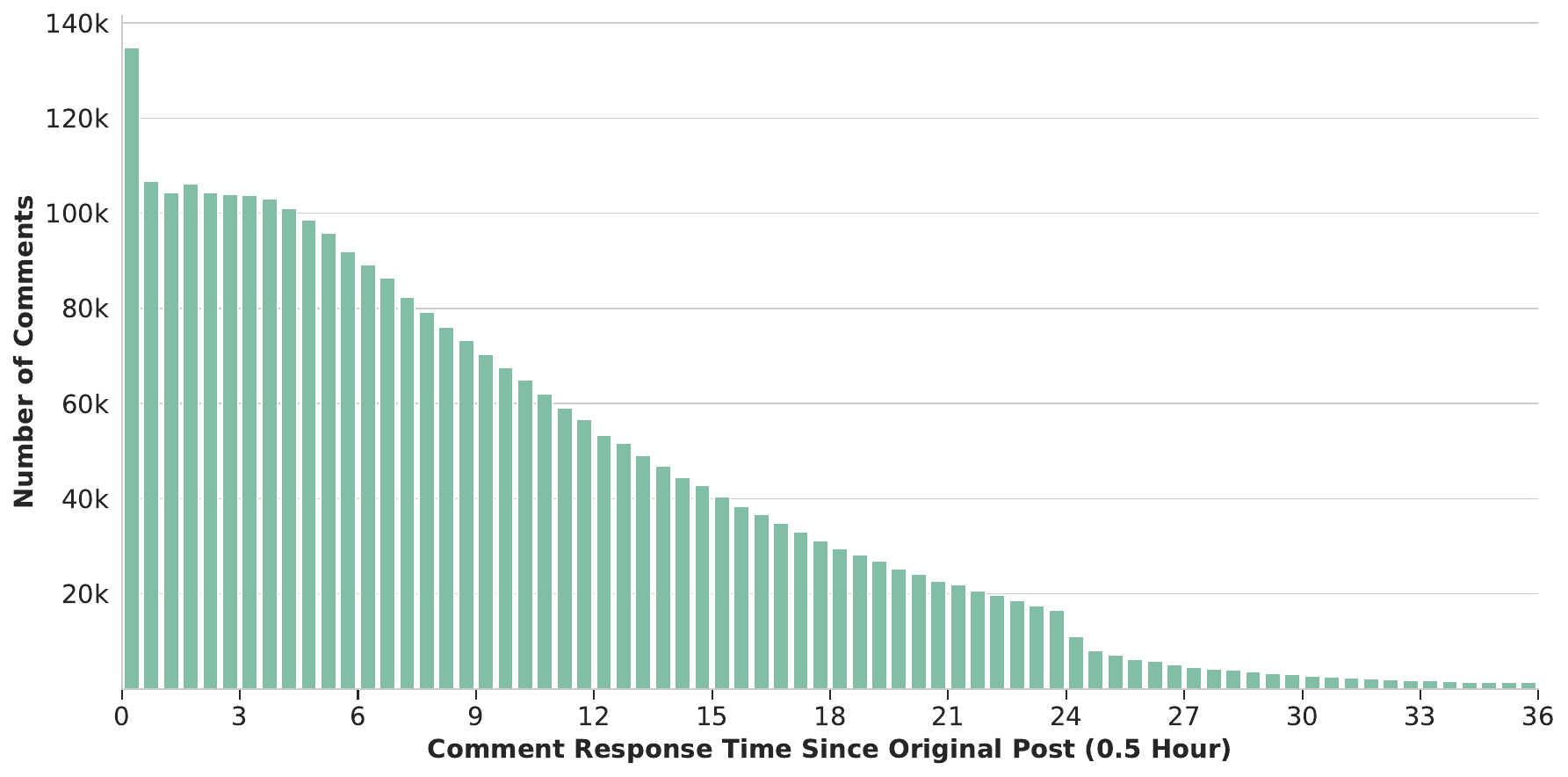}
  \caption{Distribution of comment timeliness in relation to the original post to the original post. Comment response time is binned by 30 minute intervals and labeled in hours. 4.35\% (7.80\%) of comments were posted within the first 30 minutes (1 hour) after the original post. The frequency of comments declines after this initial hour. Additionally, 93.47\% of comments were posted within the first 24 hours of the original post. Comments generated by official Reddit bots are excluded in this plot.
  }
    \label{fig:time-dist}
\end{figure}

We observe that a large proportion of comments were posted within the first 30 minutes (6.00\%) and the first hour (9.39\%) after the original post, significantly higher than the number of comments posted in the subsequent hours. This pattern is consistent with statistics reported on other platforms, such as Twitter, where 50\% of reactions occur within the first 30 minutes (\cite{wiselytics2014}), and Facebook, where the peak is within the first two hours (\cite{wisemetrics2013}).  We also note a significant number of comments posted early on may have been generated by bots, including comments directly identifiable as automatically posted by the Reddit platform. After removing the comments generated by the official Reddit bots, 4.35\% and 7.80\% of comments were posted within the first 30 minutes and 1 hour respectively. Figure~\ref{fig:time-dist} shows the volume of comments declines over time, with a sharp drop-off after the first 24 hours. This discontinuity is likely due to how Reddit displays the age of a post, where posts between 1-23 hours are labeled `x hr. ago', while posts over 24 hours are labeled `x days ago'. This finding is also consistent with where the majority of help-seeking threads in health-related forums have a life cycle of one day (\cite{wang2012stay}).  Given this skewed distribution, it is expected that the overall quality of the comments is heavily influenced by these early posts.


\begin{figure*}[htbp]
  \centering
  \includegraphics[width=\textwidth]{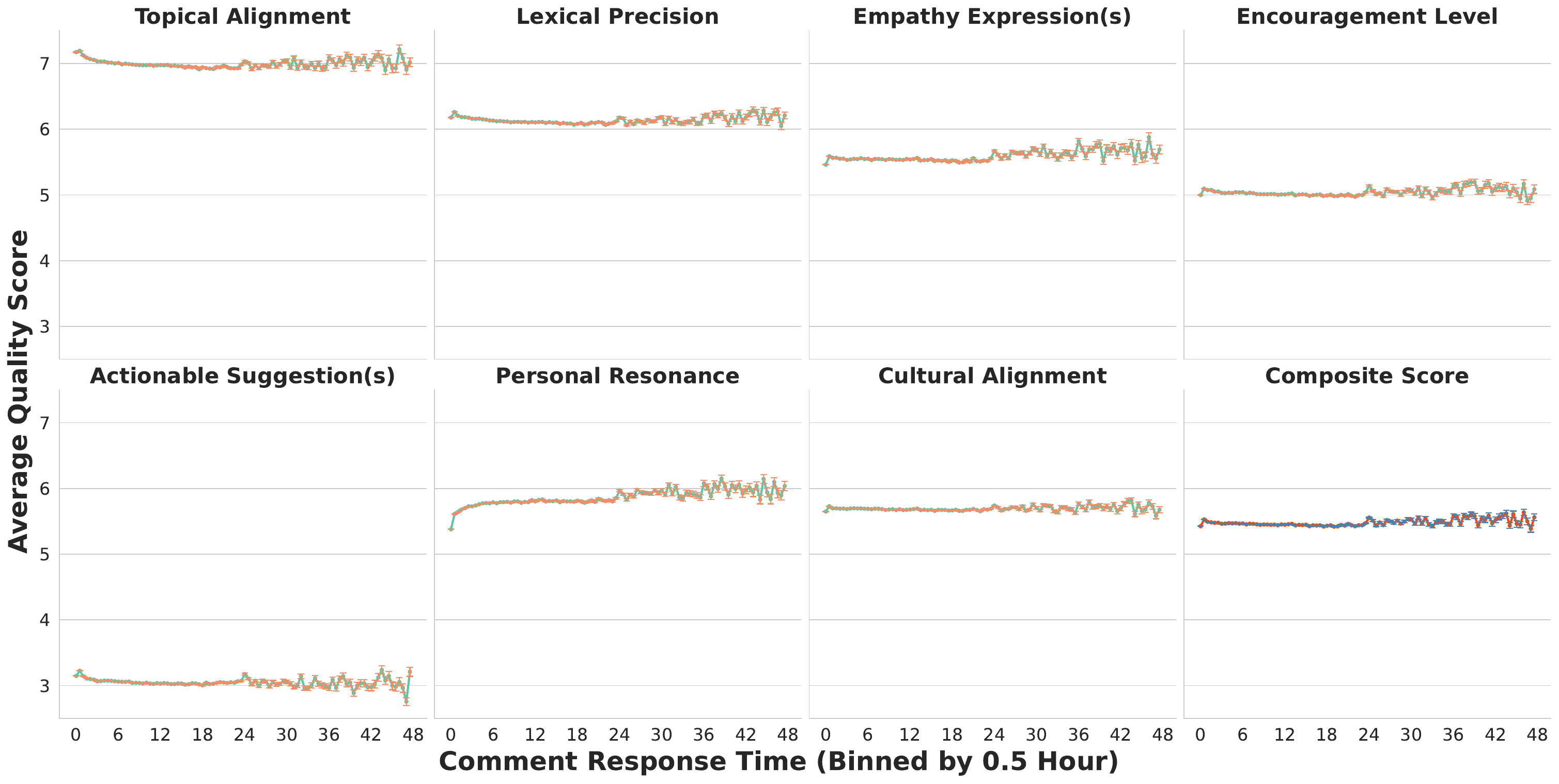}
  \caption{Average comment quality in relation to comment response time. Comment response time is binned by 30 minute intervals and labeled in hours. Comments posted within the first 30 minutes present significantly lower quality than those responding in the next 30 minutes ($p < .05$ for all eight quality measurements). The average of comment quality remains stable after the first hour, except for \textit{personal resonance} and \textit{empathy expression}, where the quality scores show a slight increase over time.}
  \label{fig:quality-over-time}
\end{figure*}

To understand the influence of early comments in the overall quality of comments responding to a post, we bin the comments by every half an hour following the corresponding posts and average the per-aspect quality scores of the comments within each bin. As shown in Figure ~\ref{fig:quality-over-time}, the overall quality of comments posted within the first half an hour is lower than those posted in the subsequent half an hour. Comments posted after the first full hour do not exhibit significant variation in quality, with the exception of \textit{personal resonance} and \textit{empathy expression}, where we observe a slight upward trend over time.


\begin{figure}[htbp]
  \centering
  \includegraphics[width=0.7\linewidth]{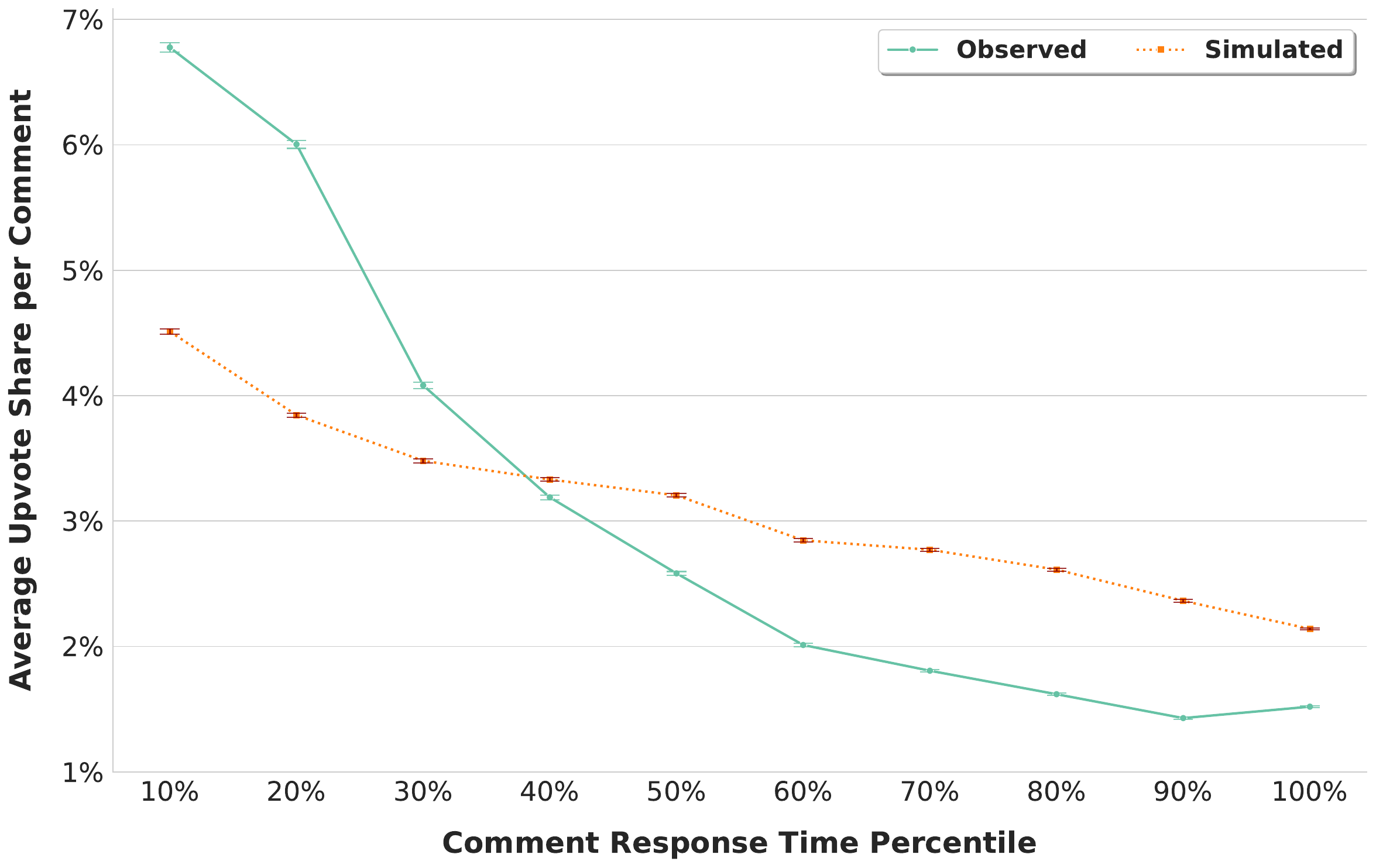}
  \caption{The first 30\% of comments on a post receive more upvotes, even after accounting for the time effect. The green line represents the average upvote share (the upvote share  refers to the number of upvotes a comment received divided by the total number of upvotes received by all comments under the same post) a comment within each decile is expected to receive. The orange line shows the average share of simulated upvotes a comment in each decile receives, assuming upvotes arrive evenly over time and are randomly distributed among available comments. The orange line does not diminish for the latest comments (100\%), as upvotes kept coming in after all comments were posted. The observed data shows a more skewed allocation of upvotes towards early comments, compared with simulated upvotes. } 
  \label{fig:upvotes-over-time}
\end{figure}

Despite the compromised quality of the earliest comments, we observe that the earlier a comment is posted, the more upvotes it tends to receive. To account for the competitive nature of upvotes among comments on the same post, we grouped comments based on their response timeliness relative to other comments. As shown in Figure~\ref{fig:upvotes-over-time}, as a comment's response time (measured by its percentile among all comments to the same post) increases, the average share of upvotes under the same post it receives decreases rapidly. This pattern may simply be a consequence of earlier comments having more visibility. To address this ``early bird'' effect, we simulated upvotes by assuming voters arrive evenly over time and randomly allocate upvotes to comments available at the time of voting. While this simulation also produced an ``earlier-gets-richer'' pattern, the effect was less pronounced than the real data. This suggests that, beyond the early bird effect, an additional factor further boosts upvotes for earlier comments. A likely explanation is that voters tend to pay more attentions to higher-ranked comments among all available, which are more likely to be the earlier one, therefore reinforcing a ``rich-gets-richer'' dynamic. 


\begin{figure}[tbp]
  \centering
  \includegraphics[width=\textwidth]{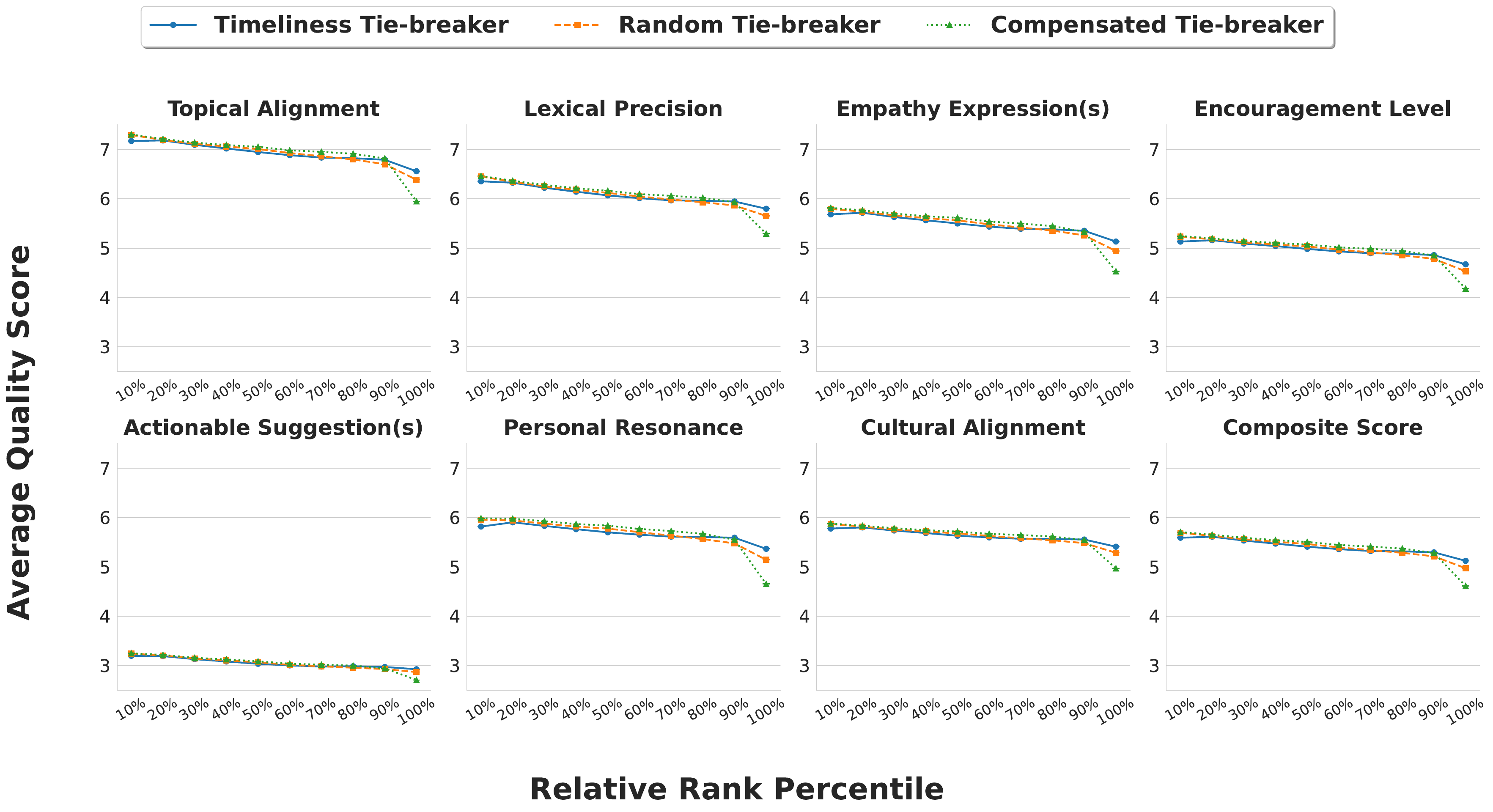}
  \caption{The top 10\% of comments, as ranked by the platform (i.e., Timeliness Tiebreaker), have lower overall ($p<.001$) and per-aspect quality in terms of \textit{emphathy expression(s)} ($p<.001$), \textit{encouragement level} ($p<.001$), \textit{personal resonance} ($p<.001$), \textit{cultural alignment} ($p<.001$), \textit{composite score} ($p<.001$) than the next decile. However, when applying a simple re-ranking strategy that randomly breaks ties for comments with the same number of upvotes (i.e., Random Tiebreaker), this pattern is reversed as the top 10\% comments outperform the second decile in all quality aspects (with all $p<.001$ except $p<.05$ for \textit{personal resonance}). Similarly, using a compensated strategy that breaks the tie by assigning a higher rank to a newer comment (i.e., Compensated Tiebreaker), comments of the top 10\% demonstrate higher quality than the second decile (with all $p<.001$), except for \textit{personal resonance} ($p=0.21$).} 
  \label{fig:quality-amended-ranking}
\end{figure}

If response time affects the quality of top-ranked comments, then controlling for this timeliness effect should improve the quality of those top-ranked comments. We illustrate this with a simple strategy. Among comments on the same post, a large proportion are tied with at least one other comment in terms of upvotes. Reddit ranks comments by upvotes, and when there is a tie, the platform breaks the tie based on the time the comment was posted—favoring the earlier comment, and granting them a lasting advantage in attracting more attention. We apply a simple adjustment by using either a random tiebreaker or a reversed tiebreaker (i.e., compensating newer comments) instead of the early-comes-first tiebreaker. Specifically, we generate a hypothetical ranking of comments based on their final upvotes as of December 2023, shuffling comments with the same number of upvotes in random order. We then re-bin the comments by rank and calculate the average quality for each bin. As shown in Figure \ref{fig:quality-amended-ranking}, this simple adjustment, regardless of random or reversed tiebreaker, raises the quality of comments ranked in the top decile above those in lower-ranked bins, reversing the pattern observed in the original data. It is surprising that even a simple adjustment to the tiebreaker can direct users' attention to higher-quality comments—an approach that platforms could easily adopt. It’s important to note that the re-ranking was applied only at the end state and did not account for the `rich-gets-richer' dynamic that unfolds throughout the ranking process. By controlling the unfair advantage of early comments in ranking, we could anticipate that high-quality content would be promoted earlier, thereby attracting more attention and engagement over time.

\section{Discussion}

The key distinction between help-seeking posts and other social media posts, such as information-sharing, opinion, or socialization posts, lies in the nuanced context embedded in the former. These posts, particularly those related to health problems, often reveal personal challenges, reflect the user’s vulnerability and uncertainty, and carry an implicit or explicit sense of urgency. As a result, the desired responses to these posts must balance several factors, including accuracy and thoughtfulness, empathy and emotional support, actionable insights, and timeliness. Our findings suggest that platforms like Reddit tend to prioritize timeliness of responses, often at the expense of other crucial dimensions such as empathy and actionability. This bias toward early responses may inadvertently direct the attention of help-seekers and readers to timely but suboptimal replies, as crafting high-quality, well-rounded responses takes time and effort. 

As noted in the literature, providing emotional support is critical to the success of health-related communities (\cite{wang2012stay}). Our analysis shows that the emotional dimensions of comments (e.g., empathy, encouragement, personal resonance) tend to score lower than topical dimensions (such as topical alignment and lexical precision). These aspects are particularly important in responding to mental health issues. While information retrieval systems are used to measuring the topical relevance of a comment to a post, there is a lack of scalable methods to assess the more nuanced emotional aspects of comments. The recent advancement of large language models offers an unprecedented accessible and scalable solution to capture these emotional nuances and provide a more holistic assessment of the usefulness of responses and unleashes the opportunity to guide the readers attentions to high-quality content. 

Additionally, most responses lack actionable suggestions. While actionable advice should ideally come from trustworthy sources, such as healthcare professionals, it is challenging to allocate their limited time to the vast volume of posts. Addressing this challenge requires finding effective ways to solicit timely, personalized, empathetic, and culturally aligned advice to better support help-seeking communities.

It is encouraging to observe that even a simple adjustment to the tiebreaker can improve the quality of top-ranked comments. This approach can be easily implemented on platforms that currently use response time as the tiebreaker. Hypothetically, the platform could adopt a more progressive approach by directly ranking comments based on their quality, an option that only became feasible with the advent of large language models. However, further research is needed to explore more sophisticated ranking strategies that promote high-quality comments as they appear, while balancing timeliness and diversity across both content and emotional dimensions. Our preliminary study highlights these possibilities and lays the groundwork for future advancements.  Additionally, it remains to be understood how different attention allocation strategies actually impact the outcomes for help-seekers and influence the behavior of content creators and consumers. 

Our research is limited to a single domain, mental health, and a single platform, Reddit, and the generalizability of our results should be tested on other platforms and domains. Beyond timeliness, there may be additional factors, external to the quality of comments, that influence their rankings. These factors might include the reputation of the responder, the social network structure, or platform-specific content promotion mechanisms. Such factors may interact with comment quality in complex ways. For example, highly specialized responders may have limited time and may not provide personalized responses, while responses that adhere strictly to community norms might overlook the cultural context of the help-seeker. Future work is needed to explore how these factors affect the quality of top-ranked comments and the distribution of user attention on the platform.

\bibliographystyle{unsrt}
\bibliography{references}

\end{document}